# Temperature-Dependence of the Resistivity of a Dilute 2D Electron System in High Parallel Magnetic Field

K. M. MERTES, HAIRONG ZHENG, S. A. VITKALOV, M. P. SARACHIK

*Physics Department, City College of the City University of New York, New York, NY 10031*

T. M. KLAPWIJK

*Department of Applied Physics, Delft University of Technology, 2628 CJ Delft, The Netherlands*

(October 27, 2018)

We report measurements of the resistance of silicon MOSFETs as a function of temperature in high parallel magnetic fields where the 2D system of electrons has been shown to be fully spin-polarized. In a field of 10.8 T, insulating behavior is found for densities up to $n_s \approx 1.35 \times 10^{11}$ cm$^{-2} \approx 1.5 n_c$; above this density the resistance is a very weak function of temperature, varying less than 10% between 0.25 K and 1.90 K. At low densities $\rho \to \infty$ more rapidly as the temperature is reduced than in zero field and the magnetoresistance $\Delta\rho/\rho$ diverges as $T \to 0$.

A great deal of interest is currently focussed on the behavior of dilute strongly interacting two-dimensional (2D) systems of electrons and holes. The resistance of these materials displays strongly insulating behavior below some critical density, $n_c$, above which metallic temperature dependence is observed, suggesting that there is a metal-insulator transition and an unexpected metallic phase in two dimensions [1]. One of the most interesting properties of these dilute 2D systems is their enormous positive magnetoresistance in reponse to magnetic fields applied parallel to the plane of the electrons. As the field is raised the resistivity increases dramatically by several orders of magnitude, the total change depending on density and temperature, and then saturates to a new field-independent plateau value above a density-dependent field $H_{sat}$ [2–6]. From an analysis of the positions of Shubnikov-de Haas oscillations in tilted magnetic fields, Okamoto et al. [7] have argued that the magnetic field above which the resistivity saturates is the field required to fully polarize the electron spins. A more direct demonstration of the onset of complete spin alignment for $H_\parallel \approx H_{sat}$ has recently been provided by small-angle Shubnikov-de Haas measurements of Vitkalov et al. [8]. Thus, the value of the resistance appears to be determined by the degree of spin polarization of the 2D electron system.

The temperature dependence of the resistance of dilute 2D systems has been measured in the absence of a magnetic field in a number of different materials. In this paper, we report measurements of the temperature dependence of the resistivity of high-mobility silicon MOSFETs in a high magnetic field of 10.8 T applied parallel to the plane of the electrons.

Measurements were made on two silicon MOSFETs of mobility approximately $20,000 V/$cm$^2$ at 4.2 K using a gate to control electron densities between $0.64 \times 10^{11}$ cm$^{-2}$ and $3.60 \times 10^{11}$ cm$^{-2}$. Contact resistances were minimized by using a split-gate geometry that allows a higher electron density in the vicinity of the contacts than in the 2D system under investigation. DC four-probe methods were used, and excitation currents were kept to a minimum to avoid heating of the electrons and insure that measurements were taken in the linear I-V regime. Data were taken in a $^3$He Oxford Heliox system for temperatures between 0.25 K and 1.90 K in magnetic fields to 10.8 T. The sample was aligned parallel to the magnetic field by using a rotating platform to minimize the transverse Hall voltage.

The resistance at $T = 0.25$ K of a silicon MOSFET is shown in the inset to Fig. 1 for various electron densities as a function of magnetic field applied parallel to the plane of the electrons. In agreement with earlier findings, the resistance increases with magnetic field and levels off to a constant value above a magnetic field, $H_{sat}$ which increases with electron density. A scaling analysis similar to that recently applied by Pudalov et al. [9] indicates that a field of 10.8 T is sufficient to reach the high-field saturated regime for electron densities up to $\approx 2.4 \times 10^{11}$ cm$^{-2}$.

The main part of Fig. 1 shows the resistance in a parallel magnetic field $H_\parallel = 10.8$ T as a function of temperature for nine different electron densities ranging from $0.92 \times 10^{11}$ cm$^{-2}$ to $2.03 \times 10^{11}$ cm$^{-2}$. For low electron densities the resistance increases with decreasing temperature, and the system is insulating, while for higher densities the resistance is nearly independent of temperature. The change from strongly-insulating to weak temperature dependence occurs at an electron density $\approx 1.35 \times 10^{11}$ cm$^{-2}$ or $\approx 1.5 n_c$ ($n_c$ is the critical density in zero field). We note that the resistance is still in the saturated regime at $H_\parallel = 10.8$ T at this density.

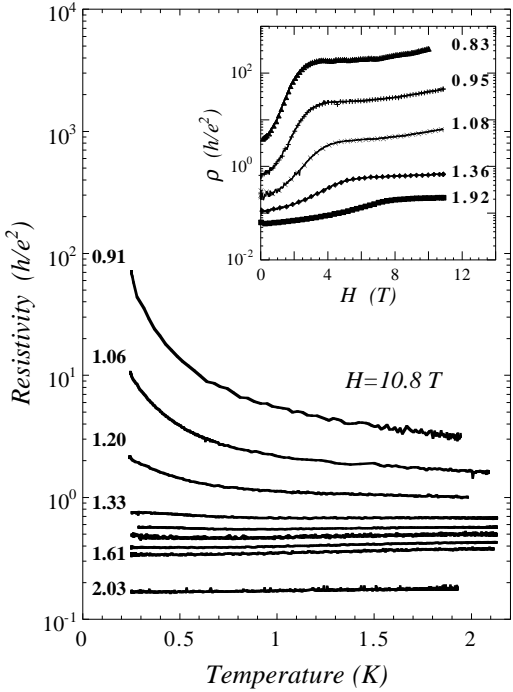

FIG. 1. Resistivity in a parallel field of 10.8 T plotted on a logarithmic scale as a function of temperature for electron densities (from top): $0.91, 1.06, 1.20, 1.33, 1.40, 1.47, 1.54, 1.61, 2.03 \times 10^{11}$ cm$^{-2}$. The critical density $n_c \approx 0.84 \times 10^{11}$ cm$^{-2}$. The inset shows the resistivity as a function of magnetic field applied parallel to the electron plane of a silicon MOSFET with different electron densities, as labelled. Except for the top curve, all densities shown in the inset are above $n_c \approx 0.84 \times 10^{11}$ cm$^{-2}$. The temperature $T = 0.25$ K.

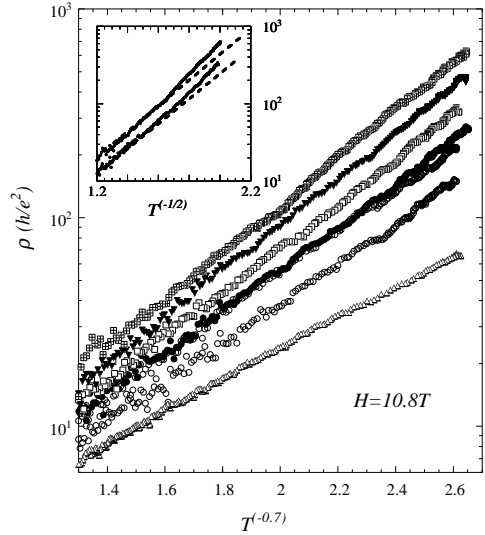

FIG. 2. Resistivity on a log scale versus $T^{-0.7}$ for electron densities (from the top): $0.817, 0.831, 0.845, 0.859, 0.873, 0.887, 0.914 \times 10^{11}$ cm$^{-2}$. The inset shows curves for $n_s = 0.817$ and $0.845 \times 10^{11}$ cm$^{-2}$ plotted as a function of $T^{-1/2}$, indicating that the data do not fit a straight line corresponding to Efros-Shklovskii variable-range hopping.

We now examine the insulating behavior in high fields in more detail. For low electron densities, Fig. 2 shows the logarithm of the resistance measured in an in-plane magnetic field of 10.8 T plotted as a function of $T^{-0.7}$. The data are consistent with straight lines; the resistance in high magnetic field can thus be approximated by the expression $\rho(H_{sat}, T) \propto \exp{(T_0/T)^{0.7}}$. In contrast, the resistance of low-density silicon MOSFETs in the absence of a magnetic field has been found to obey Efros-Shklovskii variable-range hopping, $\rho(H = 0, T) \propto \exp{(T_0/T)^{1/2}}$ [10]. To illustrate the difference, the inset to Fig. 2 shows that the high field data do not fall on straight lines when plotted as a function of $T^{-1/2}$. Instead, the resistance deviates progressively upward as the temperature is reduced (i. e., as $T^{-1/2}$ increases), indicating that it diverges more rapidly in high field than it does in zero field. Very similar behavior was found by Shlimak et al. [11] in delta-doped GaAs/AlGaAs heterostructures, where the resistance in high fields was fitted to exponentially activated variable range hopping with an exponent of approximately 0.8. As was noted by these authors, the experimentally observed resistivity can be fitted to the variable-range hopping form by increasing the value of the exponent above 1/2 or, alternatively, by using a prefactor that depends on temperature.

The temperature-dependence of the system containing equal numbers of spin-up and spin-down electrons is thus clearly different from the behavior observed when the spins are fully polarized. Although a number of explanations have been proposed to account for the strong magnetoresistance observed for electron densities on both sides of the zero-field metal-insulator transition [12–19], few of these theories have considered the effect of a magnetic field on the temperature-dependence of the resistivity. The value of the exponent $x$ in the exponentially activated hopping resistivity, $\rho \propto \exp(T_0/T)^x$, is known to depend on the form of the density of states near the Fermi energy. Specifically, a constant density of states yields Mott hopping ($x = 1/4$ in 3D, $x = 1/3$ in 2D), a soft parabolic Coulomb gap gives $x = 1/2$ in any dimension, and $x = 1$ obtains for a "hard" gap where the density of states is zero over some range of energy near $E_F$. Kurobe and Kamimura [12] pointed out that by aligning the spins of the electrons, a magnetic field supresses hops between singly-occupied states (as well as from doubly-occupied to unoccupied states). This is reflected in the shape of the density of states near the Fermi energy, with consequent changes in the exponent $x$ and the temperature dependence. This possibility could be tested through transport and simultaneous tunneling measurements to determine the single-particle density of states near $E_F$.

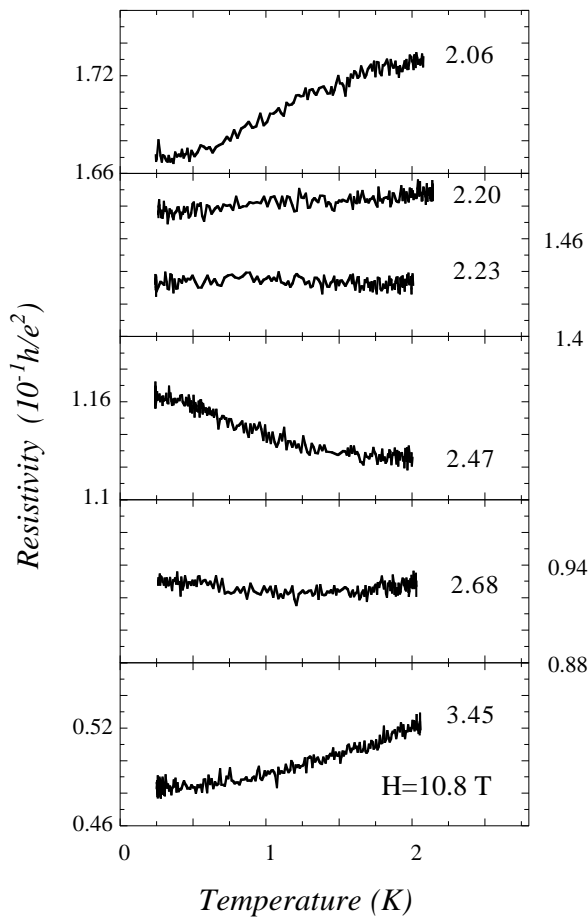

FIG. 3. Resistivity in a field of 10.8 T as a function of temperature for high electron densities, as labelled (in units $10^{11}$ cm$^{-2}$).

The magnetoresistance is larger for higher mobility samples, it is known to be bigger for lower electron densities, and it increases with decreasing temperature. Through measurements at 35 mK in high-mobility silicon MOSFETs with electron densities near the zero-field metal-insulator transition, Kravchenko and Klapwijk [20] have recently demonstrated an increase in the resistivity of more than four orders of magnitude in response to an in-plane magnetic field; the magnetoresistance is expected to be even larger at still lower temperatures. The results reported in the present paper imply that the magnetoresistance actually diverges in the limit of zero temperature. Since the resistance in high fields, $\rho(H_{sat}) \propto \exp (T_0/T)^{0.7}$, diverges more strongly than in zero field, $\rho(H = 0)) \propto \exp (T_0/T)^{1/2}$, the magnetoresistance, $[R(H_{sat}) - R(H = 0)]/R(H = 0)$, goes to infinity as $T \to 0$.

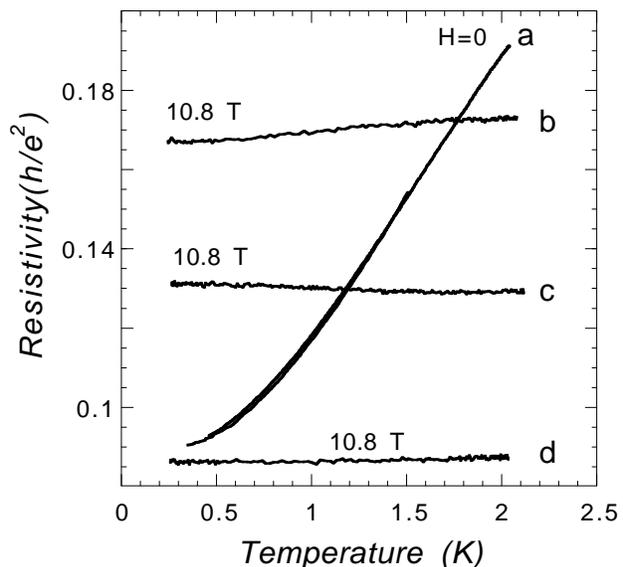

FIG. 4. Resistance versus temperature in zero field and in H=10.8 T for comparable values of resistance R. The electron densities for curves $a$ through $d$ are $1.58, 2.06, 2.34$, and $2.75 \times 10^{11}$ cm$^{-2}$.

We now briefly consider the behavior of the resistance in high fields for electron densities above $1.35 \times 10^{11}$ cm$^{-2}$. As shown in Figs. 1 and 3, the resistance for high densities depends very weakly on temperature, varying by less than 10% between 0.25K and 1.90K. As the electron density is raised, the resistance first decreases with decreasing temperature ($n_s = 2.15 \times 10^{11}$ cm$^{-2}$), then increases with decreasing temperature ($2.56 \times 10^{11}$ cm$^{-2}$), and then again decreases with decreasing temperature at very high density. These changes are all quite small, however, and the main feature is the absence of strongly localized behavior at these densities. The magnitude of the temperature dependence is comparable with that expected for quantum corrections to the Drude conductivity.

Perhaps not surprising is that localization occurs for an electron density where the resistivity is on the order of $h/e^2$. Note, however, that entry into the strongly localized regime takes place at $\rho \approx 0.6$ to $0.7h/e^2$ in high field, while the metal-insulator transition in zero field occurs at a resistivity on the order of 2 or $3h/e^2$. Using the expression for the Drude conductivity, $\rho = h/e^2 \times 2/(g_v g_s K_F l)$, (here $g_v$ and $g_s$ are the valley and the spin degeneracies, respectively) these resistivities correspond to $K_F l \approx 1.4$ to 1.7 in high field and $K_F l \approx 0.18$ to 0.25 in zero field. Although the critical resisistivity and $K_F l$ differ for different material systems (silicon versus GaAs/AlGaAs heterostructures, for example), and some variation is known to occur for a given material depending on mobility, our data are obtained for one and the same sample so that the change is solely due to the application of a parallel magnetic field. Localization is expected at $K_F l \approx 1$, which is roughly consistent with the high field behavior. In contrast, one would expect localization to be well established in zero field at $K_F l \approx 1/5$. This suggests that the process leading to localization is different in the two cases.

It is important to note that the value of the resistance does not by itself determine the size of the temperature dependence [21]. This is illustrated in Fig. 4, where the resistivity in zero field at one specific density is compared with the temperature dependence of the resistivity in high field at comparable values of resistance. In contrast with the strong metallic temperature-dependence found in the absence of a magnetic field, the resistance varies only minimally with temperature at high electron densities in high magnetic fields. Confirming earlier findings [1], our data show that a magnetic field suppresses the strong metallic behavior observed in zero field.

To summarize, we report measurements of the resistance of silicon MOSFETs for electron densities between $0.64 \times 10^{11}$ cm$^{-2}$ and $3.60 \times 10^{11}$ cm$^{-2}$ at temperatures between 0.25 K and 1.9 K in a parallel magnetic field of 10.8 T (a field sufficient to fully polarize the spins of the electrons for densities up to $2.4 \times 10^{11}$ cm$^{-2}$). For low electron densities, the resistance diverges more rapidly with decreasing temperature than it does in zero field: in contrast with Efros-Shklovskii variable-range hopping, $\rho \propto \exp(T_0/T)^x$, with exponent $x = 1/2$ found in zero field, it can be fit to the same form but with an exponent $x = 0.7$. The magnetoresistance $[R(H_{sat}) - R(H = 0)]/R(H = 0)$ thus diverges as $T \to 0$. Above $\approx 1.35 \times 10^{11}$ cm$^{-2}$, a density for which 10.8 T is well above the field necessary to fully polarize the electron spins, the resistance depends very weakly on temperature, varying less than 10% between 0.25 K and 1.90 K. The temperature dependence of the resistance in high field thus differs from its zero-field behavior for all electron densities in silicon MOSFETs.

We thank S. V. Kravchenko, Q. Si and E. Abrahams for valuable comments on the manuscript. This work was supported by DOE grant No. DOE-FG02-84-ER45153.

[1] E. Abrahams, S. V. Kravchenko, and M. P. Sarachik, to be published in Revs. Mod. Phys., preprint cond-mat/0006055 (2000); M. P. Sarachik and S. V. Kravchenko, Proc. Natl. Acad. Sci.U.S.A. **96**, 5900 (1999).
[2] V. T. Dolgopolov, G. V. Kravchenko, A. A. Shashkin, and S. V. Kravchenko, JETP Lett. **55**, 733 (1992).
[3] D. Simonian, S. V. Kravchenko, M. P. Sarachik, and V. M. Pudalov, Phys. Rev. Lett. **79**, 2304 (1997).
[4] V. M. Pudalov, G. Brunthaler, A. Prinz, and G. Bauer, JETP Lett. **65**, 932(1997).
[5] M. Y. Simmons, A. R. Hamilton, M. Pepper, E. H. Linfield, P. D. Rose, D. A. Ritchie, A. K. Savchenko, and T. G. Griffiths, Phys. Rev. Lett. **80**, 1292 (1998).
[6] J. Yoon, C. C. Li, D. Shahar, D. C. Tsui, and M. Shayegan, Phys. Rev. Lett. **84**, 4421 (2000).
[7] T. Okamoto, K. Hosoya, S. Kawaji, and A. Yagi, 1999, Phys. Rev. Lett. **82**, 3875 (1999).
[8] S. A. Vitkalov, H. Zheng, K. M. Mertes, M. P. Sarachik, and T. M. Klapwijk, preprint cond-mat/0004201 (2000).
[9] V. M. Pudalov, G. Brunthaler, A. Prinz, and G. Bauer, preprint cond-mat/0004206 (2000).
[10] W. Mason, S. V. Kravchenko, G. E. Bowker, and J. E. Furneaux, Phys. Rev. B **52**, 7857 (1995).
[11] I. Shlimak, S. I. Khondaker, M. Pepper, and D. A. Ritchie, Phys. Rev. B **61**, 7253 (2000).
[12] A. Kurobe and H. Kamimura, J. Phys. Soc of Jpn. **51**, 1904 (1982); H. Kamimura in *Electron-Electron Interaction in Disordered Systems*, edited by A. L. Efros and M. Pollak (North Holland, Amsterdam, 1985).
[13] Q. Si and C. M. Varma, Phys. Rev. Lett. **81**, 4951 (1999).
[14] S. Chakravarty, S. A. Kivelson, C. Nayak, and K. Voelker, Philos. Mag., B **79**. 859 (1999).
[15] S. Das Sarma and E. H. Hwang, Phys. Rev. Lett. **84**, 5596 (2000).
[16] V. T. Dolgopolov and A. Gold, JETP Lett. **71**, 27 (2000).
[17] T. M. Klapwijk and S. Das Sarma, Solid State Commun. **110**, 581 (1999).
[18] B. L. Altshuler and D. L. Maslov, Phys. Rev. Lett. **82**, 145 (1999).
[19] V. I. Kozub, N. V. Agrinskaya, S. I. Khondaker, and I. Shlimak, preprint cond-mat/9911450 (1999).
[20] S. V. Kravchenko and T. M. Klapwijk, Phys. Rev. Lett. **84**, 2909 (2000).
[21] This has also been explicitly demonstrated by X. G. Feng, D. Popovic, and S. Washburn, Phys. Rev. Lett. **83**, 368 (1999), as well as by S. J. Papadakis, E. P. DePoortere, and M. Shayegan, preprint cond-mat/0008041 (2000).